\documentstyle[aps,prl,floats,epsf,graphics,psfig]{revtex}

\begin{document}
\ifpreprintsty\else
\twocolumn[\hsize\textwidth%
\columnwidth\hsize\csname@twocolumnfalse\endcsname
\fi

\draft
\preprint{}

\title{Finite Temperature Behavior of the $\nu=1$ Quantum Hall Effect in Bilayer Electron Systems}

\author{M. Abolfath$^{a,b}$, Ramin Golestanian$^{c,d}$, 
and T. Jungwirth$^{e,f}$}
\address{
$^a$Department of Physics and Astronomy,
University of Oklahoma, Norman, Oklahoma 73019-0225\\
$^b$Institute for Studies in Theoretical Physics and Mathematics,
P.O. Box 19395-5531, Tehran, Iran \\
$^c$Institute for Theoretical Physics, University of California,
Santa Barbara, CA 93106-4030 \\
$^d$Institute for Advanced Studies in Basic Sciences, 
Zanjan 45195-159, Iran \\
$^e$Department of Physics, Indiana University, Bloomington, 
Indiana 47405, \\  
$^e$Institute of Physics ASCR,
Cukrovarnick\'a 10, 162 00 Praha 6, Czech Republic
}

\date{\today}

\maketitle

\begin{abstract}
\leftskip 2cm
\rightskip 2cm
An effective field theoretic description of $\nu=1$ bilayer electron systems stabilized by Coulomb 
repulsion in a single wide quantum well is examined using renormalization group techniques. 
The system is found to undergo a crossover from a low temperature strongly correlated quantum
Hall state to a high temperature compressible state. This picture is used to account for the recent
experimental observation of an anomalous transition in bilayer electron systems
(T. S. Lay, {\em et al.} Phys. Rev. B {\bf 50}, 17725 (1994)). 
An estimate for the crossover temperature is provided, and it is shown 
that its dependence on electron density is in reasonable agreement with the experiment.
\end{abstract}


\ifpreprintsty\else\vskip1pc]\fi
\narrowtext

\section{Introduction and Summary}	\label{sIntro}

The novel structure of bilayer electron systems
(BLESs) has recently garnered considerable attention
\cite{Lay94,Jungwirth,DasSarma,Mac1,Kyang94,Wen,Moon,Kyang96,Abolfath}.
These systems have proven to be  
appropriate candidates for probing quantum phases of 
electron systems, such as the quantum Hall effect (QHE).
A BLES can be made in a double quantum well, where
a high and hard-wall barrier leads to formation of two separate
layers of electrons. It may also be realized in a single wide quantum 
well (SWQW), in which  
the barrier separating the electron layers originates from Coulomb repulsion
of electrons in the well \cite{Abolfath}.

The low energy effective theory for a BLES can be described as an
XY spin model with an in-plane magnetic field, in which the spin-ordered 
phase corresponds to a correlated quantum Hall state\cite{Mac1,Moon,Kyang96}. 
In this spin analogy, local spin orientations encode the coherence 
between the two layers, the spin stiffness is a measure of loss of exchange 
and correlation energy corresponding to spatial variation of the relative
coherence, and the magnetic field is given by the  
tunneling amplitude. In the absence of the tunneling term, 
the system exhibits a finite temperature Kosterlitz-Thouless 
(KT) phase transition \cite{KT}, which destroys the phase coherence
and thus the QHE \cite{Kyang94,Wen,Moon}. 
For any non-zero 
(but small) value of the tunneling parameter, however, the transition is 
known to be smoothed out into a crossover, from a low temperature saturated 
ferromagnetic regime to a high temperature disordered regime \cite{Wen,Kyang96,Jose,Nelson}.

Recently, Lay {\em et al.} have reported an experimental observation
of a finite temperature quantum Hall ``phase transition''
in a BLES, realized in a SWQW \cite{Lay94}.  
They find that collapse of the QHE gap takes place at a temperature
$T^\ast$, whose value {\em decreases} when the electron areal density $N_s$ is {\em increased}.


Here we study the effect of thermal fluctuations on the quantum Hall
behavior of a BLES in a SWQW. We sketch the derivation of the effective field
theory for this system 
and argue that an additional Ising-like symmetry-breaking term should be added
to the Hamiltonian. We provide numerical estimates for  the coupling constants 
of the theory using  Hartree-Fock (HF)
variational wavefunction of the many-particle system combined 
with the self-consistent local spin density approximation (LSDA) for single particle states.  
We then study finite temperature behavior of the system using renormalization group 
(RG) techniques. We argue that the system undergoes a crossover from a low temperature
ordered regime to a high temperature disordered regime and that the experimentally observed
$N_s$-dependence of $T^{\ast}$ is related to the monotonically decreasing {\em iso-spin}
stiffness with increasing electron density in the SWQW.

\section{Effective Hamiltonian} 	\label{sEff}

The low energy state of a BLES at $\nu=1$ can be well described
by the following variational wave function \cite{Moon,Kyang96}
\begin{equation}
   |\Psi\rangle = \prod_X \left(\hat{c}^\dagger_{X \uparrow}
      +e^{i\phi(X)} \hat{c}^\dagger_{X \downarrow} \right)|0\rangle,
\label{wave}
\end{equation}
where the up or down chiral iso-spin states denote localized electrons
in the left or right side of the well, respectively, and $X$ is a quantum number
such as the Landau gauge orbital guiding center.
The phase angle field $\phi(X)$ denotes the relative local coherence
between the two electron layers, and is well known to entail lowest
energy excitations of the BLES \cite{Moon,Kyang96}.
In Eq.(1) we have neglected the real-spin degree of freedom of the BLES. This 
approximation is justified by the self-consistent
LSDA calculations yielding a unique, fully
spin-polarized  solution at $\nu=1$ for the whole range of electron densities used in 
the experiment \cite{Lay94}. The calculated Zeeman splitting, 
$\Delta_z$, is strongly enhanced by the exchange-correlation energy and always 
larger than the gap between two lowest energy levels
in the SWQW, as shown in Table~\ref{Tab1}. 

The general form of the effective Hamiltonian for a BLES in a SWQW
has been obtained in Ref. \cite{Abolfath}, where it has been assumed that the BLES can
be described effectively by the Hilbert space of the two lowest subbands.
The effective energy functional can be written as ${\cal H}_{\rm eff}=V_{\rm HF}+T$, 
where $T$ is the tunneling term and
the microscopic HF Coulomb energy of the BLES 
$V_{\rm HF} \equiv \langle\Psi| V |\Psi\rangle$ is obtained as
\begin{eqnarray}
V_{\rm HF} &=& -\frac{1}{4} \sum_{X_1,X_2} 
V_{X_1,X_2,X_2,X_1}^{\uparrow\downarrow\uparrow\downarrow}
[m_x(X_1) m_x(X_2)
\nonumber \\&&
+ m_y(X_1) m_y(X_2)]
+ V_{X_1,X_2,X_2,X_1}^{\uparrow\uparrow\downarrow\downarrow}
\nonumber \\&&  \times
[m_x(X_1) m_x(X_2)
- m_y(X_1) m_y(X_2)],
\label{HF}
\end{eqnarray}
with ${\bf m}(X) = (\cos\phi(X), \sin\phi(X))$. 
The Coulomb energy has been evaluated
using the Landau gauge, 
where
\begin{eqnarray}
V^{\sigma_1\sigma_2\sigma_3\sigma_4}_{X_1, X_2, X_2, X_1} =
\int \frac{d^2 {\bf q}}{(2\pi)^2} &&
V^{\sigma_1\sigma_2\sigma_3\sigma_4}({\bf q}) e^{-q^2\ell^2_0/2}
\nonumber \\&& \times
\delta\left(\frac{X_1 - X_2}{\ell^2_0} - q_y\right),
\label{8.0}
\end{eqnarray}
and
\begin{eqnarray}
V^{\sigma_{1}\sigma_{2}\sigma_{3}\sigma_{4}}({\bf q}) &=&
\frac{2\pi e^2}{\epsilon q} \int dz_{1}\int dz_{2}
\psi^{\ast\sigma_{1}}(z_{1})\psi^{\ast\sigma_{2}}(z_{2})
\nonumber \\&& \times
\psi^{\sigma_{3}}(z_{1})
\psi^{\sigma_{4}}(z_{2}) e^{-q|z_{1}-z_{2}|},
\label{8.1}     
\end{eqnarray}
are the appropriate form factors,
$\epsilon$ is the dielectric constant of the host semiconductor,
and $\ell_0=\sqrt{\hbar c/e B}$ is the magnetic length.

In the continuum limit, the effective energy functional
has a gradient expansion that reads
\begin{eqnarray}
{\cal H}_{\rm eff} &=& \int d^2 {\bf r} 
\left[\frac{\rho_s}{2} (\nabla \phi)^2
-\frac{t}{2\pi\ell_0^2}\cos\phi
-\frac{\kappa^2}{2\pi\ell_0^2} \cos2\phi \right],
\label{eff}
\end{eqnarray}
in which we have (only) neglected higher derivative terms.
The origin of the iso-spin stiffness
$\rho_s$, and $\kappa^2$ that corresponds to ``pair hopping'',
is the HF loss of exchange energy 
\begin{eqnarray}
\rho_s &=& \frac{\ell_0^2}{32\pi^2} \int dq \; q^3 \left(
V^{\uparrow\downarrow\uparrow\downarrow}(q) -
V^{\uparrow\uparrow\downarrow\downarrow}(q) \right)
e^{-q^2\ell_0^2/2}, \nonumber \\
\kappa^2 &=& \frac{1}{8\pi} \int dq \;
q V^{\uparrow\uparrow\downarrow\downarrow}(q) e^{-q^2\ell_0^2/2}.
\label{coup}
\end{eqnarray}
and $t$ is the tunneling amplitude, defined as one-half of the spacing between
two lowest energy levels in the SWQW.
For the BLESs with large enough overlap integral in the middle of the
well $V^{\uparrow\uparrow\downarrow\downarrow}$, and thus $\kappa^2$, 
are significant. Values of the couplings $\rho_s$, $t$ and $\kappa^2$, of the
layer separation $d$, and of the Zeeman splitting calculated for the sample used by
Lay {\em et al.} \cite{Lay94} 
are presented in Table \ref{Tab1}.

\begin{table}
\begin{tabular}{lcccccc}
$N_{s}$ &
$\rho_s $(K) &
$t$ (K) &
$\kappa^2$ (K) &
$d/\ell_0$ &
$\Delta_z$ (K) 
\\ \tableline
0.8   & 0.15 & 6.4  & 0.19 & 3.0 & 20.2 \\
1.0   & 0.12 & 5.1  & 0.15 & 3.7 & 22.7 \\
1.2   & 0.10 & 4.1  & 0.12 & 4.2 & 25.1 \\
1.4   & 0.08 & 3.4  & 0.10 & 4.7 & 27.4 \\
1.6   & 0.06 & 2.9  & 0.07 & 5.2 & 29.6 \\
\end{tabular}
\vskip0.3cm
\caption{The SWQW parameters for various densities $N_s (10^{11} {\rm cm}^{-2})$
at $\nu=1$, and given well width of 750${\rm \AA}$ corresponding to the sample
used by Lay {\em et al.} \protect\cite{Lay94}, obtained from Eq.(\ref{coup}) 
and self-consistent LSDA one-particle states.
}
\label{Tab1}
\end{table}


The first term in Eq.(\ref{eff}) represents the usual, rotationally invariant
superfluid exchange coupling, which yields a spontaneous phase coherent state 
\cite{Wen,Kyang96}. (Note that the field $\phi$ is compact.) 
The above Hamiltonian is thus an XY-model with two symmetry-breaking terms:
the second and third terms in Eq.(\ref{eff}) represent a uniform in-plane
magnetic field and an Ising-like anisotropy, respectively.
In the absence of the tunneling and the Ising-like terms, 
the system exhibits a finite temperature KT
phase transition \cite{KT}, which destroys the phase coherence
and thus the QHE \cite{Wen,Moon}. 

\section{Renormalization Group Approach}	\label{sRG}

\begin{figure}
\center
\epsfxsize 5.3cm \rotatebox{-90}{\epsffile{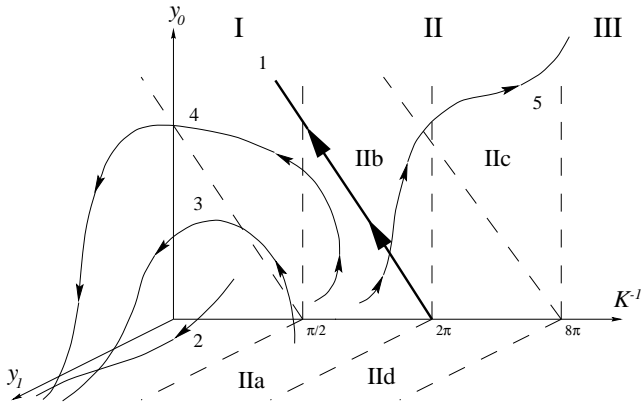}}
\vskip1cm
\caption{RG flows in the different domains of the parameter space.}
\label{Fig1}
\end{figure}

To study the XY model with the symmetry breaking terms, we follow closely
the approach by Jos\'e {\em et al.}\cite{Jose}. 
Finite temperature behavior of the system is described by the partition function
\begin{equation}
{\cal Z}= \int_{0 \leq \phi < 2\pi} {\cal D} \phi  ~e^{-{\cal H}_{\rm eff}/k_B T}.
\end{equation}
The essence of the symmetry-breaking terms can be 
captured by performing Villain expansions, which introduce discrete Coulomb-like charge 
species, called type-1 and type-2 vortices \cite{Jose,Nelson}. On the other hand,
the compactness of the field $\phi$ is accounted for by introducing yet another 
type of vortex, called type-0 vortex, which describes singular 
behavior of an otherwise noncompact field $\phi'$; a combination that serves as a substitute 
for $\phi$ \cite{Nelson,SGT}. The latter vortex introduces a new coupling constant,
namely the fugacity $y_0=e^{-E_c/k_B T}$,
which is controlled by the core energy of a type-0 vortex $E_c=2 \pi \rho_s e_c$, where
$e_c$ is a numerical constant close to 1 \cite{ec=1}. 
The corresponding three-species coupled Coulomb gas can then be studied using standard 
renormalization group (RG) techniques \cite{Nelson}. The RG flow equations for the dimensionless
coupling constant $K=\rho_s/k_B T$, and the ``fugacities'' $y_0$, $y_1=t/8 \pi^2 \rho_s$, and 
$y_2=\kappa^2/4 \pi^2 \rho_s$ are obtained as

\begin{eqnarray}
\frac{d K^{-1}}{dl} &=& 4\pi^3 \left(y_0^2 -y_1^2-y_2^2 \right),
\nonumber \\
\frac{dy_0}{dl} &=& \left(2 - \pi K \right) y_0,
\nonumber \\
\frac{dy_1}{dl} &=& \left(2 - \frac{1}{4\pi K}\right) y_1,
\nonumber \\
\frac{dy_2}{dl} &=& \left(2 - \frac{1}{\pi K}\right) y_2,
\label{RGE}
\end{eqnarray}
to the leading order in the fugacities. 
A scaling argument shows that the coupling $y_2$ is always subleading
compared to $y_1$. We thus set it to zero for the moment to
simplify the RG picture of the problem, and will comment on the effects
due to a nonzero $y_2$ later. The RG flow diagram for Eq.(\ref{RGE})
is shown in Fig.~\ref{Fig1}.

The system described by Eq.(\ref{eff}) (for $\kappa^2=0$) is known to
have a duality in the parameter domain where the Villain expansion
is applicable. The duality maps the high temperature regions to the low
temperature ones, with the roles of the fugacities being exchanged
and temperature being inverted\cite{Jose,Nelson}. This necessitates
the existence of {\em self-dual} points, {\em i.e.} those which
remain invariant under the duality transformation, at some intermediate
temperature. It is easy to see from Eq.(\ref{RGE}) that 
$K^{-1}=2\pi$ and $y_0=y_1$ define the line of self-dual points,
which is denoted as path 1 in Fig.~\ref{Fig1}.
We note that in Fig.~\ref{Fig1}, we have only sketched the RG flow 
structure for $K^{-1}< 2\pi$, and that the corresponding
behaviors in the $K^{-1}>2\pi$ region can be understood using 
this duality. 

The above RG equations show that there are three different regions 
in the parameter space: (I) $0< K^{-1}< \pi/2$, in which $y_0$ is 
irrelevant and $y_1$ is relevant, (II) $\pi/2 < K^{-1} < 8\pi$, in 
which both $y_0$ and $y_1$ are relevant, and (III) $K^{-1} > 8 \pi$,
in which $y_0$ is relevant and $y_1$ is irrelevant. 
In region I, the symmetry breaking term is dominant, as illustrated 
by path 2 in Fig.~\ref{Fig1}, and the system
is in a locked-in regime. While type-0 vortices are bound in pairs,
type-1 vortices are unbound. On the other hand, region III (that is dual 
to region I) is a disordered regime, in which type-0 vortices 
are unbound and type-1 vortices are bound in pairs. 

Region II consists of four different areas. 
Path 3 shows a typical behavior for region IIa defined by
$\pi/2<K^{-1}<2\pi$, and $y_1>y_0$.
While $K^{-1}$ is monotonically decreasing under the RG flow, both
$y_0$ and $y_1$ initially tend to increase with $y_1$ being always
exponentially dominant compared to $y_0$, until the flow eventually
passes onto region I. Although there are always some unbound type-0
vortices, the overwhelmingly larger number of free type-1 vortices 
ensures that the system is still in the locked-in regime.

Region IIb, which is defined by $\pi/2<K^{-1}<2\pi$, and $y_1<y_0$, 
is a crossover domain. During the early stages of the RG flow, 
both $y_0$ and $y_1$ increase to approach the $y_0=y_1$ surface, while
$K^{-1}$ is also increasing. The paths seem to 
be approaching the self-dual line asymptotically as if they are
being attracted to an ``intermediate temperature fixed point.''
However, at some point they will eventually
exit the region, and find their ways to either the low temperature
(such as path 4) or the high temperature (such as path 5) regions.
The fact that the renormalized fugacities approach the $y_0=y_1$ 
surface means that there are about as many unbound type-0 and 
type-1 vortices, which explains why this is a crossover region. 
Finally, the behavior of regions IIc ($2 \pi<K^{-1}<8\pi$, and  
$y_1<y_0$) and IId ($2 \pi<K^{-1}<8\pi$, and  $y_1>y_0$), 
can be understood using the duality.

For an XY model in a magnetic field, a crossover from a low
temperature locked-in regime to a high temperature disordered
regime has indeed been suggested in the literature by many authors 
\cite{Wen,Kyang96,Jose,Nelson}. However, it is not clear how this 
picture can be reconciled with the experimental observation by 
Lay {\em et al.}, which is suggestive of a finite temperature 
phase transition in the same system \cite{Lay94}.
To resolve this issue, we note that the presence of the 
self-dual line, and the fact that some RG flows in the crossover
region are seemingly attracted to it, makes the crossover
so rapid that it might ``look like'' a phase transition
that is smeared out due to finite size effects.

To make an estimate for the temperature at which this 
{\it rapid crossover} takes place, it is necessary
to characterize the crossover more carefully \cite{comment}.
Since in region II both fugacities are relevant, we
need to worry about the validity of the perturbative
RG approach. Each RG equation describing exponential
growth of a fugacity presumably breaks down at a length 
scale $\xi$ at which the fugacity is renormalized to unity. 
This correlation length simply characterizes the average 
separation between unbound vortices of the corresponding 
type, and is a measure of the number of free vortices 
\cite{Nelson}. Therefore, it is reasonable to assume that
the crossover happens where the correlation length for type-0 
vortices equals that of type-1 vortices.

Putting this together with the above picture for a 
rapid crossover, we assert that the crossover 
corresponds to an initial point, given by the bare parameters
$K^{-1}(0)$, $y_0(0)$, and $y_1(0)$, which is going to be
renormalized to $K^{-1}(l^\ast)\simeq 2\pi$, and
$y_0(l^\ast)\simeq y_1(l^\ast)\simeq 1$ via the RG flow, where
$l^\ast=\ln(\xi/a)$, and $a$ is a microscopic length scale
set by the core radius of a vortex.
This criterion yields two equations for the two
unknown parameters $\xi$ and the crossover temperature $T^\ast$,
which could in principle be solved numerically.

\begin{figure}
\center
\epsfxsize 6.3cm \rotatebox{-90}{\epsffile{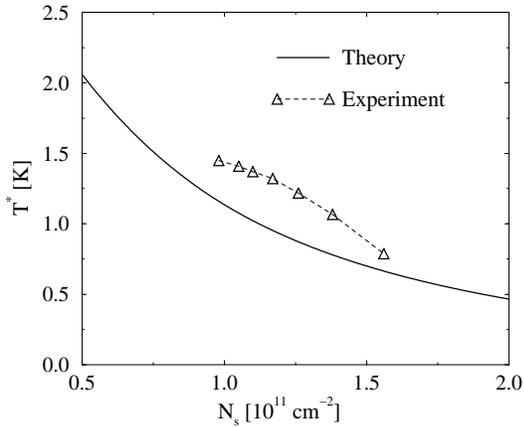}}
\vskip0.5cm
\caption{Comparison of the theoretical (solid line) 
and experimental (triangles) crossover temperatures as a function of
the BLES density.}
\label{Fig0}
\end{figure}

Rather than elaborating on the numerical calculations
that are not particularly illuminating, we attempt to make further 
simplifying approximations. We roughly estimate the integrals as
$\int_0^{l^\ast} d t (1/2-\pi K(t)) \simeq c_0 l^\ast (1/2 -\pi K(0))$
and
$\int_0^{l^\ast} d t (1/2-1/4\pi K(t)) \simeq c_1 l^\ast (1/2-1/4\pi K(0))$,
where the coefficients $0< c_0 <1$ and $0< c_1 <1$ describe the decay 
of the coupling constant $K^{-1}$. Using this approximation, we obtain
an estimate for the crossover temperature as
\begin{equation}
T^\ast\simeq{2\pi \rho_s \over k_B}
\times \left[\frac{c_0 \ln(8\pi^2 \rho_s/t)+(3+c_1) e_c}
{(3+c_0) \ln(8\pi^2 \rho_s/t)+c_1 e_c}\right]. \label{T*}
\end{equation}
Values for $T^{\ast}$, given by Eq.(\ref{T*}) with $c_0=c_1=0.3$ and
$e_c=1$, are shown in Fig.~\ref{Fig0} together with experimental data from
Ref.~\cite{Lay94}. The magnitude of the crossover temperature is of the order 
of 1~K, and it decreases with increasing the BLES density, in reasonalbe
agreement with the experiment.

\section{Discussion}	\label{sDiscuss}

A nonzero value for $\kappa^2$ does not change the RG picture, because
the Ising-like term is always subdominant compared to the magnetic
field, except for the low temperature regime where the behavior of 
the system is governed by the ground state properties. While
the Hamiltonian (\ref{eff}) has only one minimum
for $t/\kappa^2>4$, a double-well structure with
a local minimum at $\phi=\pi$, a global minimum at $\phi=0$, and a maximum at
$\phi = \cos^{-1}(-t/4\kappa^2)$ develops for $t/\kappa^2 \leq 4$.
If the system is initially prepared to be at the metastable state,
it will decay to the ground state with a rate 
$\sim \exp(-S[\phi_c] / \hbar)$
where $\phi_c$ is the classical solution,
interpolating between the two minima \cite{Coleman}.
As one may see from Table \ref{Tab1}, the condition for the existence
of the metastable state ($\kappa^2 \geq t/4$) is not satisfied 
for a typical BLES. 
However, it might be possible to reduce the value of $t$
by applying a tilted magnetic field\cite{Jungwirth}.

Finally, we note that a full quantitative account of the experiment
can not be achieved until one correctly takes into account all the details,
possibly by performing an exact diagonalization calculation.
However, we believe our effective field theory approach to the problem,
provides an understanding of the nature of the observed ``transition'' in the 
experiment, and is able to predict the correct order of magnitude and trend
for the ``transition'' temperature.

\acknowledgments

We are grateful to B. Davoodi and M.R. Rahimi-Tabar who
were involved in the early stages of this study and to T.S. Lay and
M. Shayegan for discussions and for providing with experimental
data.
It is a pleasure to thank H. Fertig, M.P.A. Fisher, S.M. Girvin, 
A. H. MacDonald, S. Rouhani, and A. Zee for invaluable 
discussions. 
MA acknowledges support from EPS-9720651 and a grant from the Oklahoma
State Regents for Higher Education.
The work at ITP was supported by the NSF under Grant 
No. PHY94-07194. TJ acknowledges support by NSF under grant DGE-9902579
and by the Grant Agency of the Czech Republic
under grant 202/98/0085.



\begin{references}


\bibitem{Lay94}
T. S. Lay, Y. W. Suen, H. C. Manoharan, X. Ying, M. B. Santos, and M. Shayegan,
Phys. Rev. B {\bf 50}, 17725 (1994).

\bibitem{Jungwirth}
T. S. Lay, T. Jungwirth, L. Smr\v{c}ka, and M. Shayegan,
Phys. Rev. B {\bf 56}, R7092 (1997).

\bibitem{DasSarma}
J. P. Eisenstein, 
in {\em Novel Quantum Liquids in Semiconductor Structures}, edited by
S. DasSarma and A. Pinczuk (Wiley, New York, 1997).


\bibitem{Mac1}
A. H. MacDonald, P. M. Platzman, and G. S. Boebinger, 
Phys. Rev. Lett. {\bf 65}, 775 (1990);
L. Brey, Phys. Rev. B {\bf 65}, 903 (1990);
H. Fertig, Phys. Rev. B {\bf 40}, 1087 (1989).

\bibitem{Kyang94}
K. Yang, K. Moon, L. Zheng, A. H. MacDonald, S. M. Girvin,
D. Yoshioka, and S. -C. Zhang, Phys. Rev. Lett. {\bf 72}, 732 (1994).


\bibitem{Wen}
X. -G. Wen, A. Zee, Phys. Rev. Lett. {\bf 69}, 1811 (1992);
Phys. Rev. B {\bf 46}, 2290 (1992); Phys. Rev. B {\bf 47}, 2265 (1993).

\bibitem{Moon}
K. Moon, H. Mori, K. Yang, S. M. Girvin, A. H. MacDonald,
L. Zheng, D. Yoshioka, S. Zhang, Phys. Rev. B {\bf 51}, 5138 (1995).
S. M. Girvin, A. H. MacDonald, in {\em Novel Quantum Liquids
in Semiconductor Structures}, edited by S. DasSarma and
A. Pinczuk (Wiley, New York, 1996).

\bibitem{Kyang96}
K. Yang, K. Moon, L. Belkhir, H. Mori, S. M. Girvin, and
A. H. MacDonald, L. Zheng, and D. Yoshioka, 
Phys. Rev. B {\bf 54}, 11644 (1996).


\bibitem{Abolfath}
M. Abolfath, L. Belkhir, and N. Nafari, 
Phys. Rev. B {\bf 55}, 10643 (1997).

\bibitem{KT}
J.~M.~Kosterlitz and D.~J.~Thouless, J. Phys. C {\bf 6}, 1181 (1973);
V.~L.~Berezinskii, Zh. Eksp. Teor. Fiz. {\bf 61}, 1144 (1971) [Sov. Phys.
JETP {\bf 34}, 610 (1972)].

\bibitem{Jose}
J. Jos\'e, L. P. Kadanoff, S. Kirkpatrick, and D. R. Nelson, \prb {\bf 16},
1217 (1977).

\bibitem{Nelson}
D. R. Nelson, in {\em Phase Transitions and Critical Phenomena} Vol. 7, edited
by C. Domb and J. L. Lebowitz (Academic Press, London, 1983).

\bibitem{SGT}
Due to compactness of the field, this model is {\it not} a 
sine-Gordon field theory, unless the self-energy of the vortices becomes
very large that fluctuations around the boundarys of $\phi$ becomes unlikely
($\phi \approx \phi'$) and the KT phase transition occurs at $T_{KT} = 8\pi\rho_s/k_B$.
In this case, $y_0 = 0$ is the sine-Gordon plane.

\bibitem{ec=1}
M. Abolfath, J. J. Palacios, H. A. Fertig, S. M. Girvin, and A. H. MacDonald, 
Phys. Rev. B {\bf 56}, 6795 (1997).


\bibitem{comment}
It was in fact suggested by Lay {et al.} \cite{Lay94} 
that a rapid crossover is not inconsistent with their observation.


\bibitem{sensitive}
We note that the corresponding trend is neither sensitive to the
approximations involved in the derivation of Eq.(\ref{T*}), nor
to the choice of numerical coefficients within the range of validity
of the perturbative approach.

\bibitem{Coleman}
J. S. Langer, Ann. Phys. {\bf 54}, 258 (1969).





\end{references}
\end{document}